\documentclass[
aps,
prd,
reprint,
superscriptaddress,
nofootinbib,
amsmath,
amssymb,
longbibliography
]{revtex4-2}

\usepackage[utf8]{inputenc}
\usepackage[T1]{fontenc}
\usepackage{lmodern}
\usepackage{graphicx}
\usepackage{dcolumn}
\usepackage{bm}
\usepackage{mathtools}
\usepackage{xcolor}
\usepackage[colorlinks=true,linkcolor=blue,citecolor=blue,urlcolor=blue]{hyperref}

\newcommand{\diag}{\operatorname{diag}}
\newcommand{\ren}{\mathrm{ren}}

\begin{document}

\title{Semiclassical Backreaction of Massive Quantum Fields in the Spacetime of a Global Monopole}

\author{Owen Pavel Fernández Piedra}
\email{opavelfp2006@gmail.com}
\affiliation{Department of Physics, Center for Exact and Natural Sciences, Federal University of Paraíba, 58059-970, João Pessoa, Brazil}

\date{\today}

\begin{abstract}
We study the leading semiclassical backreaction of massive scalar, spinor,
and vector fields in the spacetime of a pointlike global monopole. Starting
from the renormalized stress-energy tensors obtained in the
Schwinger--DeWitt approximation, we derive the first-order semiclassical
geometry generated by a general conserved diagonal source and then
specialize it to fields of spin $0$, $1/2$, and $1$. The locally fixed
quantum source falls as $r^{-6}$ and induces metric corrections of order
$r^{-4}$. The scalar field generically produces distinct temporal and radial
metric functions, whereas the spinor and vector fields lead to one-function
geometries. We analyze the resulting curvature correction, the acceleration
of static observers, geodesic motion, and the validity domain of the
first-order solution. The quantum backreaction produces a local deformation
of the monopole exterior but does not modify its asymptotic solid-angle
deficit.
\end{abstract}

\maketitle

\section{Introduction}
\label{sec:introduction}

Topological defects arise naturally in field theories with spontaneous
symmetry breaking. Their existence is determined by the topology of the
vacuum manifold, and their gravitational fields provide simple but nontrivial
laboratories for studying the relation between matter, geometry, and quantum
effects in curved spacetime
\cite{Kibble1976,VilenkinShellard1994,HindmarshKibble1995}. A particularly
useful example is the global monopole, associated with the symmetry-breaking
pattern $O(3)\longrightarrow O(2)$.

Outside the monopole core, the scalar order parameter approaches the vacuum
manifold, but its angular gradients do not vanish. The exterior energy density
therefore falls as $r^{-2}$, and the total energy contained inside a sphere
grows linearly with the radial cutoff. In the pointlike approximation, where
the detailed structure of the core is not resolved, the corresponding exterior
geometry is the Barriola--Vilenkin spacetime
\cite{BarriolaVilenkin1989,HarariLousto1990,BezerraDeMello2001}. For
vanishing Schwarzschild-like mass parameter, the metric can be written as
\begin{equation}
ds_0^2
=
-\alpha^2 dt^2
+
\frac{dr^2}{\alpha^2}
+
r^2 d\Omega^2 ,
\label{eq:intro_background_metric}
\end{equation}
with
\begin{equation}
\alpha^2=1-8\pi\eta^2 ,
\label{eq:intro_alpha}
\end{equation}
in units $G=c=\hbar=1$, where $\eta$ is the symmetry-breaking scale. The
parameter $\alpha$ encodes the angular deficit of the geometry. Indeed, after
the coordinate redefinitions $\tau=\alpha t$ and $\rho=r/\alpha$,
Eq.~\eqref{eq:intro_background_metric} becomes
\begin{equation}
ds_0^2
=
-d\tau^2
+
d\rho^2
+
\alpha^2\rho^2d\Omega^2 .
\label{eq:intro_rescaled_metric}
\end{equation}
Thus, a sphere of proper radial coordinate $\rho$ has area
$4\pi\alpha^2\rho^2$, and the solid-angle deficit is
\begin{equation}
\Delta\Omega=4\pi(1-\alpha^2).
\label{eq:intro_solid_angle_deficit}
\end{equation}
The spacetime is therefore not asymptotically flat in the usual sense, even
though its local curvature becomes small at large radius.

Quantum field theory in curved spacetime provides a semiclassical framework
in which the geometry is treated classically while the matter fields are
quantized \cite{BirrellDavies1982,Fulling1989,Wald1994,ParkerToms2009}. The
central local quantity is the renormalized stress-energy tensor
$\left\langle T^\mu{}_\nu\right\rangle_{\ren}$, which appears as a source in
the semiclassical Einstein equations
\cite{Wald1977,HuVerdaguer2008,Simon1991,ParkerSimon1993}
\begin{equation}
G^\mu{}_\nu[g_{\mu\nu}]
=
8\pi
\left(
T^\mu{}_\nu
+
\left\langle T^\mu{}_\nu\right\rangle_{\ren}
\right),
\label{eq:intro_semiclassical_equations}
\end{equation}
where $T^\mu{}_\nu$ is the classical stress-energy tensor of the monopole.
The renormalization of the stress-energy tensor can be formulated through
covariant point-splitting, local momentum-space methods, or equivalent local
and covariant prescriptions
\cite{Christensen1976,Christensen1978,BunchParker1979,
HollandsWald2001,Moretti2003,DecaniniFolacci2008}.

In the present work, the unperturbed classical monopole metric
$g^{(0)}_{\mu\nu}$ satisfies
\begin{equation}
G^\mu{}_\nu[g^{(0)}_{\mu\nu}]
=
8\pi T^\mu{}_\nu .
\label{eq:intro_background_equations}
\end{equation}
The backreacted metric is written perturbatively as
\begin{equation}
g_{\mu\nu}
=
g^{(0)}_{\mu\nu}
+
h_{\mu\nu},
\qquad
\left|h_{\mu\nu}\right|
\ll
\left|g^{(0)}_{\mu\nu}\right| .
\label{eq:intro_metric_perturbation}
\end{equation}
Expanding Eq.~\eqref{eq:intro_semiclassical_equations} to first order around
the classical geometry gives
\begin{equation}
\delta G^\mu{}_\nu[h_{\mu\nu}]
=
8\pi
\left\langle T^\mu{}_\nu\right\rangle_{\ren}^{(0)},
\label{eq:intro_linearized_equations}
\end{equation}
where the superscript $(0)$ indicates that the renormalized stress-energy
tensor is evaluated on the unperturbed monopole background. Corrections to
the quantum source induced by $h_{\mu\nu}$ contribute only beyond the
first-order approximation considered here.

For massive fields, the Schwinger--DeWitt expansion provides a local
large-mass approximation to the renormalized effective action and to the
corresponding stress-energy tensor
\cite{DeWitt1965,BarvinskyVilkovisky1985,Avramidi1991,
Avramidi1998Erratum,Avramidi2000,Vassilevich2003,OttewillWardell2011}. The
approximation is controlled by the ratio between the Compton wavelength of
the quantized field and the local curvature radius. In the monopole spacetime,
the curvature scale is of order
\begin{equation}
L_{\rm curv}
\sim
\frac{r}{\sqrt{1-\alpha^2}},
\end{equation}
so for a quantum field of mass $\mu$ the large-mass condition is
\begin{equation}
\mu r \gg \sqrt{1-\alpha^2}.
\label{eq:intro_sdw_condition}
\end{equation}
In this regime, the leading nontrivial contribution to
$\left\langle T^\mu{}_\nu\right\rangle_{\ren}$ is of order $\mu^{-2}$.
Applications of this approximation to static spherically symmetric and
black-hole spacetimes have a long history
\cite{FrolovZelnikov1982,AndersonHiscockSamuel1995,Matyjasek2000,
MatyjasekTrynieckiZwierzchowska2010}.

Quantum effects in global-monopole geometries have previously been studied
for scalar and spinor fields, both in the idealized pointlike spacetime and
in models with a finite core
\cite{Hiscock1990,MazzitelliLousto1991,BezerraMelloBezerraKhusnutdinov1999,
BezerraMelloSaharian2007}. In particular, the renormalized stress-energy
tensors for massive scalar, spinor, and vector fields in the pointlike
global-monopole spacetime have been obtained in the Schwinger--DeWitt
approximation
\cite{FernandezPiedra2019ScalarSET,FernandezPiedra2020Spinor,
FernandezPiedra2020Vector}. Related results for the scalar vacuum
polarization in the same geometry also allow one to examine the structure and
accuracy of higher-order local terms in the large-mass expansion
\cite{FernandezPiedra2020Phi2}.

The purpose of this paper is to determine the first-order semiclassical
metric produced by those quantum sources. The calculation is carried out in
two steps. First, we solve the semiclassical Einstein equations for a
conserved diagonal source with general coefficients. Then, we insert the
scalar, spinor, and vector coefficients obtained from the corresponding
renormalized stress-energy tensors. This separates the part of the
backreaction that follows from the conserved structure of the source from the
part that depends on the field spin, mass, and, for the scalar field, the
curvature coupling.

We also analyze the physical content of the corrected geometries. In
particular, we compute the correction to the Ricci scalar, determine the
acceleration required for static observers to remain at fixed radius, and
study the effective potential and radial kinetic factor for timelike and null
geodesics. These quantities show explicitly that the semiclassical source
does not modify the asymptotic deficit angle, but generates a local
deformation of the exterior geometry.

The paper is organized as follows. In Sec.~\ref{sec:classical_monopole} we
review the classical pointlike global-monopole geometry and fix the
conventions. In Sec.~\ref{sec:renormalized_tensor} we summarize the
renormalized stress-energy tensors used as quantum sources. In
Sec.~\ref{sec:general_backreaction} we solve the semiclassical Einstein
equations for a general conserved $r^{-6}$ source.
Sections~\ref{sec:scalar_backreaction}, \ref{sec:spinor_backreaction}, and
\ref{sec:vector_backreaction} present the scalar, spinor, and vector
geometries. In Sec.~\ref{sec:comparison_spins} we compare the three
backreactions. In Sec.~\ref{sec:physical_analysis} we discuss curvature
corrections, static acceleration, geodesic motion, and the regime of
validity. Finally, Sec.~\ref{sec:conclusions} summarizes the results.

\section{Classical global monopole geometry}
\label{sec:classical_monopole}

In  this section we will briefly recall the classical global monopole geometry in order to fix the conventions used throughout the paper \cite{BarriolaVilenkin1989,HarariLousto1990}. The simplest model is described by a triplet of real scalar fields $\Phi^a$, $a=1,2,3$, with
\begin{equation}
\mathcal L_\Phi
=
-\frac{1}{2}g^{\mu\nu}
\partial_\mu\Phi^a\partial_\nu\Phi^a
-
\frac{\lambda}{4}
\left(
\Phi^a\Phi^a-\eta^2
\right)^2 .
\label{eq:monopole_lagrangian}
\end{equation}
The vacuum manifold is defined by $\Phi^a\Phi^a=\eta^2$ and is therefore isomorphic to $S^2$. Since $\pi_2(S^2)=\mathbb Z$, the theory admits monopole configurations. For a static global monopole one takes the hedgehog ansatz
\begin{equation}
\Phi^a
=
\eta h(r)\frac{x^a}{r},
\label{eq:hedgehog_ansatz}
\end{equation}
with $h(r)\to1$ outside the core. In the exterior region, $r\gg r_{\rm c}$, the potential energy is negligible, but the angular gradient energy remains. The corresponding stress-energy tensor behaves as
\begin{equation}
T^t{}_t
\simeq
T^r{}_r
\simeq
-\frac{\eta^2}{r^2},
\qquad
T^\theta{}_{\theta}
=
T^\varphi{}_{\varphi}
\simeq
0.
\label{eq:monopole_exterior_stress_tensor}
\end{equation}
Thus the monopole is not a localized source in the usual sense. Its energy density falls as $r^{-2}$, so the total energy inside a sphere grows linearly with the radius. This long-range stress-energy distribution is responsible for the solid-angle-deficit character of the geometry.

Using the static spherically symmetric form
\begin{equation}
ds^2
=
-A(r)\,dt^2
+
\frac{dr^2}{B(r)}
+
r^2d\Omega^2,
\label{eq:monopole_metric_ansatz}
\end{equation}
the $tt$ Einstein equation involves only the radial metric function $B(r)$. After integrating it we obtain
\begin{equation}
B(r)
=
\alpha^2-\frac{2M}{r},
\label{eq:monopole_radial_metric_function}
\end{equation}
where
\begin{equation}
\alpha^2=1-8\pi\eta^2.
\label{eq:alpha_classical_section}
\end{equation}
Here, $M$ is an integration constant that is not determined by the exterior stress-energy tensor alone. It depends on the physics of the monopole core and plays the role of a Schwarzschild-like mass parameter.

The difference between the radial and temporal Einstein equations is
\begin{equation}
G^r{}_r-G^t{}_t
=
\frac{B(r)}{r}
\frac{d}{dr}
\ln\!\left(
\frac{A(r)}{B(r)}
\right).
\label{eq:monopole_temporal_radial_difference}
\end{equation}
Because the exterior source satisfies $T^r{}_r=T^t{}_t$, the corresponding difference on the matter side vanishes. Equation~\eqref{eq:monopole_temporal_radial_difference} then implies that $A(r)/B(r)$ is constant. This constant can be absorbed by a constant rescaling of the time coordinate, so that $A(r)=B(r)$. The exterior geometry is consequently
\begin{equation}
ds^2
=
-\left(
\alpha^2-\frac{2M}{r}
\right)dt^2
+
\left(
\alpha^2-\frac{2M}{r}
\right)^{-1}dr^2
+
r^2d\Omega^2 .
\label{eq:monopole_with_mass}
\end{equation}

Since the present work focuses on the idealized pointlike monopole background, we set the core-dependent parameter $M$ to zero. The unperturbed geometry is therefore obtained from Eq.~\eqref{eq:monopole_with_mass} with $M=0$, which gives $A_0(r)=B_0(r)=\alpha^2$, resulting in Eq.~\eqref{eq:intro_background_metric}. Thus setting $M=0$ does not remove the monopole source; it only removes the Schwarzschild-like contribution associated with the unresolved core.

The parameter $\alpha$ also has a direct geometrical interpretation. Consider the spatial geometry on a hypersurface of constant $t$. For the pointlike background, the induced line element is
\begin{equation}
d\ell_0^2
=
\frac{dr^2}{\alpha^2}
+
r^2d\Omega^2.
\label{eq:classical_spatial_line_element}
\end{equation}
At fixed $r$, the angular part of this metric is
$d\sigma^2=r^2d\Omega^2$. Therefore, $r$ is an areal radius, i.e., the symmetry sphere labelled by $r$ has area $\mathcal A(r)=4\pi r^2$.
The coordinate $r$ thus determines the size of the angular two-spheres. It does not, however, measure the physical radial separation between two such spheres.

To obtain that separation, consider a radial curve within the constant-$t$ hypersurface. Along this curve the angular coordinates remain fixed, so that Equation~\eqref{eq:classical_spatial_line_element} implies that the infinitesimal physical radial distance associated with a coordinate increment $dr$ is $d\ell_0=dr/\alpha$.
Choosing a reference sphere at $r=r_\star$, with $r_\star$ lying in the exterior region and $\ell_0(r_\star)=0$, the proper radial distance from this reference sphere to a sphere of areal radius $r$ is
\begin{equation}
\ell_0(r)
=
\int_{r_\star}^{r}d\ell_0
=
\int_{r_\star}^{r}\frac{dr'}{\alpha}
=
\frac{r-r_\star}{\alpha}.
\label{eq:classical_proper_radius}
\end{equation}
The use of $r_\star$ is appropriate because the pointlike metric is employed only outside the monopole core. Changing $r_\star$ changes $\ell_0(r)$ only by an additive constant and therefore cannot modify the large-radius behavior.

Now solving Equation~\eqref{eq:classical_proper_radius} for the areal radius gives $r=\alpha\ell_0+r_\star$, so that a symmetry sphere situated at large proper radial distance $\ell_0\gg r_\star$ from the reference sphere has area $\mathcal A(\ell_0)
=
4\pi\alpha^2\ell_0^2
+
\mathcal O(\ell_0)$.
The coefficient of the leading $\ell_0^2$ term is therefore the solid angle $\Omega_\infty
=
4\pi\alpha^2$, so that the solid-angle deficit relative to the Euclidean value $4\pi$ is
\begin{equation}
\Delta\Omega_\infty
=
4\pi-\Omega_\infty
=
4\pi\left(1-\alpha^2\right).
\label{eq:classical_solid_angle_deficit}
\end{equation}
Thus, the pointlike global monopole has an asymptotically conical spatial geometry. 

Finally, for the pointlike global monopole the Ricci scalar is
\begin{equation}
R_0
=
\frac{2(1-\alpha^2)}{r^2}.
\label{eq:R0}
\end{equation}
The solution will be used only in the exterior region
$r\gg r_{\rm c}$, where $r_{\rm c}$ is the characteristic radius of the monopole core. This condition is part of the pointlike approximation: the radial coordinate is assumed to be large compared with the size of the core, so the internal profile of the scalar field is not resolved.

\section{Renormalized stress-energy tensors}
\label{sec:renormalized_tensor}

We now specify the renormalized stress-energy tensors in the pointlike global monopole background that provide the quantum sources for the semiclassical Einstein equations, and where previously obtained in references \cite{FernandezPiedra2019ScalarSET,FernandezPiedra2020Spinor,FernandezPiedra2020Vector}.

For compactness, we write $\Delta_\alpha\equiv1-\alpha^2$. The results quoted below are the leading terms of the Schwinger--DeWitt large-mass expansion. Thus, for a quantized field of spin $s$ and mass $\mu_s$, they are applicable when $\mu_s r\gg\sqrt{\Delta_\alpha}$, which is equivalent to requiring that the Compton wavelength of the field be much smaller than the local curvature radius, $L_{\rm curv}\sim r/\sqrt{\Delta_\alpha}$. In this regime, the renormalized stress-energy tensor is local and its leading contribution is of order $\mu_s^{-2}$\cite{BunchParker1979,BarvinskyVilkovisky1985,Vassilevich2003}.

At the leading order considered here, the renormalized stress-energy tensors of the scalar, spinor, and vector fields have the common static, spherically symmetric diagonal form
\begin{equation}
\left\langle T^\mu{}_\nu\right\rangle_{\ren}^{(0)}
=
\frac{1}{r^6}
\diag
\left(
C_t,
C_r,
C_\perp,
C_\perp
\right),
\label{eq:general_diagonal_quantum_tensor}
\end{equation}
where $C_\perp$ is the common coefficient of the $\theta\theta$ and $\varphi\varphi$ components, whose equality follows from spherical symmetry. Covariant conservation further relates this angular coefficient to the radial one, so that $C_\perp$ is not independent. The radial component of covariant conservation gives
\begin{align*}
\nabla_\mu
\left\langle T^\mu{}_r\right\rangle_{\ren}^{(0)}
=
\frac{-6C_r+2(C_r-C_\perp)}{r^7}=0,
\end{align*}
and therefore $C_\perp=-2C_r$, which gives for the conserved source
\begin{equation}
\left\langle T^\mu{}_\nu\right\rangle_{\ren}^{(0)}
=
\frac{1}{r^6}
\diag
\left(
C_t,
C_r,
-2C_r,
-2C_r
\right).
\label{eq:conserved_quantum_tensor}
\end{equation}
For the massive scalar field, the coefficients are
\begin{align}
C_t^{\rm sc}
&=
-\frac{\Delta_\alpha}{10080\pi^2\mu_{\rm sc}^2}
\mathcal B(\alpha,\xi),
\nonumber\\
C_r^{\rm sc}
&=
\frac{\Delta_\alpha}{10080\pi^2\mu_{\rm sc}^2}
\mathcal Q(\alpha,\xi),
\label{eq:scalar_coefficients}
\end{align}
where $\xi$ is the curvature coupling, and the functions $\mathcal B(\alpha,\xi)$ and $\mathcal Q(\alpha,\xi)$ are cubic polynomials in $\xi$ given by
\begin{equation}
\begin{aligned}
\mathcal B(\alpha,\xi)
&=
B_0(\alpha)
+
B_1(\alpha)\xi
+
B_2(\alpha)\xi^2
+
B_3(\alpha)\xi^3,
\\
\mathcal Q(\alpha,\xi)
&=
Q_0(\alpha)
+
Q_1(\alpha)\xi
+
Q_2(\alpha)\xi^2
+
Q_3(\alpha)\xi^3.
\end{aligned}
\label{eq:scalar_BQ_polynomials}
\end{equation}
with
\begin{equation}
\begin{aligned}
B_0(\alpha)
&=
101\alpha^2(1+\alpha^2)-4,
\\
B_1(\alpha)
&=
-504\alpha^4-1554\alpha^2+42,
\\
B_2(\alpha)
&=
-3150\alpha^4+8400\alpha^2-210,
\\
B_3(\alpha)
&=
15540\alpha^4-15960\alpha^2+420.
\end{aligned}
\label{eq:scalar_B_coefficients}
\end{equation}
and
\begin{equation}
\begin{aligned}
Q_0(\alpha)
&=
4+67\alpha^2(1+\alpha^2),
\\
Q_1(\alpha)
&=
-336\alpha^4+966\alpha^2-42,
\\
Q_2(\alpha)
&=
-1890\alpha^4+5040\alpha^2+210,
\\
Q_3(\alpha)
&=
-9660\alpha^4+9240\alpha^2-420.
\end{aligned}
\label{eq:scalar_Q_coefficients}
\end{equation}
For a generic curvature coupling, $C_t^{\rm sc}\neq C_r^{\rm sc}$.

For a massive spinor field of mass $\mu_{\rm sp}$, the temporal and radial coefficients coincide, $C_t=C_r\equiv C_{\rm sp}$, with
\begin{equation}
C_{\rm sp}
=
\frac{
\Delta_\alpha
\left(
31\alpha^4+31\alpha^2+10
\right)
}{
40320\pi^2\mu_{\rm sp}^2
}.
\label{eq:spinor_coefficient}
\end{equation}

Likewise, for a massive vector field of mass $\mu_{\rm v}$, one has $C_t=C_r\equiv C_{\rm v}$, with
\begin{equation}
C_{\rm v}
=
\frac{
\Delta_\alpha
\left(
25\alpha^4+25\alpha^2+4
\right)
}{
3360\pi^2\mu_{\rm v}^2
}.
\label{eq:vector_coefficient}
\end{equation}
Thus the spinor and vector sources have equal temporal and radial coefficients, whereas the scalar source generically does not.

\section{General semiclassical backreaction}
\label{sec:general_backreaction}

Having specified the conserved quantum sources, we now determine the corresponding first-order backreaction for generic temporal and radial coefficients $C_t$ and $C_r$ in Eq.~\eqref{eq:conserved_quantum_tensor}. The dependence on the spin of the quantized field is therefore kept implicit and will enter only through the particular coefficients substituted into the general solution.

Although we work in units $G=c=\hbar=1$, the renormalized stress-energy tensor represents the leading one-loop contribution and would carry an explicit factor of $\hbar$ if this constant were restored \cite{BirrellDavies1982,ParkerToms2009,Wald1977}. To make the perturbative expansion explicit, we introduce a dimensionless parameter $\varepsilon$. This parameter has no independent physical meaning; it is used only to identify the order of the quantum correction. Accordingly, in Eq.~\eqref{eq:intro_semiclassical_equations} we replace $\left\langle T^\mu{}_\nu\right\rangle_{\ren}^{(0)}$ by $\varepsilon\left\langle T^\mu{}_\nu\right\rangle_{\ren}^{(0)}$ and retain terms through first order in $\varepsilon$. Since the metric perturbation is itself of order $\varepsilon$, the variation of the renormalized stress-energy tensor induced by that perturbation is also of order $\varepsilon$. Once multiplied by the explicit factor of $\varepsilon$, it contributes only at order $\varepsilon^2$ and is consistently neglected. At the end of the calculation, we set $\varepsilon=1$.

We use the static, spherically symmetric line element introduced in Eq.~\eqref{eq:monopole_metric_ansatz}. For the pointlike monopole background, the two metric functions coincide and satisfy $A_0(r)=B_0(r)=\alpha^2$. We therefore write
\begin{equation}
A(r)
=
\alpha^2
+
\varepsilon A_q(r),
\qquad
B(r)
=
\alpha^2
+
\varepsilon B_q(r),
\label{eq:perturbative_expansion}
\end{equation}
where the subscript $q$ identifies the first-order quantum correction.

For the metric of Eq.~\eqref{eq:monopole_metric_ansatz}, the temporal and radial mixed components of the Einstein tensor are
\begin{align*}
G^t{}_t
&=
\frac{B(r)-1+rB'(r)}{r^2},
\\
G^r{}_r
&=
\frac{B(r)-1}{r^2}
+
\frac{B(r)A'(r)}{rA(r)}.
\end{align*}
The classical monopole contribution satisfies $8\pi T^t{}_t=8\pi T^r{}_r=-\Delta_\alpha/r^2$. After including the quantum source with its explicit factor of $\varepsilon$ and using Eq.~\eqref{eq:perturbative_expansion}, the temporal and radial Einstein equations become
\begin{align*}
-\frac{\Delta_\alpha}{r^2}
+
\varepsilon
\frac{B_q(r)+rB_q'(r)}{r^2}
&=
-\frac{\Delta_\alpha}{r^2}
+
\frac{8\pi\varepsilon C_t}{r^6},
\\
-\frac{\Delta_\alpha}{r^2}
+
\varepsilon
\left[
\frac{B_q(r)}{r^2}
+
\frac{A_q'(r)}{r}
\right]
&=
-\frac{\Delta_\alpha}{r^2}
+
\frac{8\pi\varepsilon C_r}{r^6}.
\end{align*}

The temporal equation gives
\[
\frac{d}{dr}
\left[
rB_q(r)
\right]
=
\frac{8\pi C_t}{r^4},
\]
which integrates to
\begin{equation}
B_q(r)
=
-\frac{2M_q}{r}
-
\frac{8\pi C_t}{3r^4}.
\label{eq:Bq_solution}
\end{equation}
Here $M_q$ is an integration constant associated with the first-order correction to the Schwarzschild-like mass parameter. As will be the case for all field spins considered in this work, its value cannot be determined by the local exterior equations alone and would require boundary conditions or matching to a finite monopole core \cite{BezerraMelloSaharian2007}. In the subsequent sections, we will isolate the locally fixed vacuum-polarization effects by focusing strictly on the $r^{-4}$ terms.

Subtracting the temporal equation from the radial one isolates the difference between the two metric corrections:
\[
A_q'(r)-B_q'(r)
=
\frac{8\pi\left(C_r-C_t\right)}{r^5}.
\]
Integrating gives
\[
A_q(r)-B_q(r)
=
K_q
-
\frac{2\pi\left(C_r-C_t\right)}{r^4},
\]
where $K_q$ is a constant. This constant corresponds to a constant rescaling of the time coordinate. Choosing the time normalization such that the corrected metric approaches the pointlike monopole background at large radius, $A(r)\to\alpha^2$ and $B(r)\to\alpha^2$ as $r\to\infty$, fixes $K_q=0$. Hence,
\[
A_q(r)
=
B_q(r)
-
\frac{2\pi\left(C_r-C_t\right)}{r^4}.
\]

Substituting Eq.~\eqref{eq:Bq_solution} into this relation, and then setting $\varepsilon=1$, gives
\begin{align}
A(r)
&=
\alpha^2
-
\frac{2M_q}{r}
-
\frac{2\pi}{r^4}
\left(
C_r+\frac{C_t}{3}
\right),
\label{eq:A_general_solution}
\\
B(r)
&=
\alpha^2
-
\frac{2M_q}{r}
-
\frac{8\pi C_t}{3r^4}.
\label{eq:B_general_solution}
\end{align}

The angular Einstein equations provide an independent consistency check. In terms of the metric functions $A(r)$ and $B(r)$, they are
\begin{equation}
\begin{aligned}
G^\theta{}_\theta
=
G^\varphi{}_\varphi
&=
\frac{B(r)}{2}
\left[
\frac{A''(r)}{A(r)}
+
\frac{A'(r)}{rA(r)}
-
\frac{1}{2}
\left(
\frac{A'(r)}{A(r)}
\right)^2
\right]
\\
&\quad
+
\frac{B'(r)}{2}
\left[
\frac{A'(r)}{2A(r)}
+
\frac{1}{r}
\right].
\label{eq:angular_Einstein_components}
\end{aligned}
\end{equation}
Using Eqs.~\eqref{eq:A_general_solution} and \eqref{eq:B_general_solution}, and retaining the first-order backreaction, one finds
\[
G^\theta{}_\theta
=
G^\varphi{}_\varphi
=
-\frac{16\pi C_r}{r^6},
\]
which coincides with the $\theta\theta$ and $\varphi\varphi$ components of the source in Eq.~\eqref{eq:conserved_quantum_tensor}.

Equations~\eqref{eq:A_general_solution} and \eqref{eq:B_general_solution} give the general first-order backreacted geometry generated by the conserved quantum source. The two metric functions coincide precisely when $C_t=C_r$. In that case, the corrected metric can be written in terms of a single radial function.

\section{Scalar-field backreaction}
\label{sec:scalar_backreaction}

We now specialize the general solution to a massive scalar field \cite{FernandezPiedra2019ScalarSET}. The corresponding coefficients $C_t^{\rm sc}$ and $C_r^{\rm sc}$, together with the functions $\mathcal B(\alpha,\xi)$ and $\mathcal Q(\alpha,\xi)$, were introduced in Sec.~\ref{sec:renormalized_tensor}. To make the local corrections to the temporal and radial metric functions explicit, we define
\begin{equation}
\begin{aligned}
\gamma_{A,{\rm sc}}
&\equiv
\frac{\Delta_\alpha}{5040\pi\mu_{\rm sc}^2}
\left[
\mathcal Q(\alpha,\xi)
-
\frac{1}{3}\mathcal B(\alpha,\xi)
\right],
\\
\gamma_{B,{\rm sc}}
&\equiv
\frac{\Delta_\alpha}{3780\pi\mu_{\rm sc}^2}
\mathcal B(\alpha,\xi).
\end{aligned}
\label{eq:scalar_gamma_definitions}
\end{equation}
Substituting $C_t=C_t^{\rm sc}$ and $C_r=C_r^{\rm sc}$ into Eqs.~\eqref{eq:A_general_solution} and \eqref{eq:B_general_solution}, and denoting the corresponding integration constant by $M_{\rm q,sc}$, gives
\begin{align}
A_{\rm sc}(r)
&=
\alpha^2
-
\frac{2M_{\rm q,sc}}{r}
-
\frac{\gamma_{A,{\rm sc}}}{r^4},
\label{eq:scalar_A_function}
\\
B_{\rm sc}(r)
&=
\alpha^2
-
\frac{2M_{\rm q,sc}}{r}
+
\frac{\gamma_{B,{\rm sc}}}{r^4}.
\label{eq:scalar_B_function}
\end{align}
Together with the line element \eqref{eq:monopole_metric_ansatz}, these functions define the scalar-field backreacted geometry.

The term proportional to $M_{\rm q,sc}/r$ has a different origin from the terms proportional to $r^{-4}$. It arises as the homogeneous integration term obtained when the temporal semiclassical Einstein equation is integrated to determine the radial metric function. Consequently, the local exterior stress-energy tensor fixes $\gamma_{A,{\rm sc}}$ and $\gamma_{B,{\rm sc}}$, but does not determine $M_{\rm q,sc}$ itself. Its value must be selected by boundary conditions or by matching the exterior solution to a regular monopole core. Although it has the same radial dependence as a Schwarzschild-type mass term, $M_{\rm q,sc}/r$ should therefore not be identified with the locally determined vacuum-polarization correction.

For minimal coupling, $\xi=0$, the coefficients in Eq.~\eqref{eq:scalar_gamma_definitions} reduce to
\begin{equation}
\begin{aligned}
\gamma_{A,{\rm sc}}^{\rm m}
&=
\frac{
\left(
25\alpha^4+25\alpha^2+4
\right)\Delta_\alpha
}{
3780\pi\mu_{\rm sc}^2
},
\\
\gamma_{B,{\rm sc}}^{\rm m}
&=
\frac{
\left(
101\alpha^4+101\alpha^2-4
\right)\Delta_\alpha
}{
3780\pi\mu_{\rm sc}^2
},
\end{aligned}
\label{eq:scalar_minimal_gamma_definitions}
\end{equation}
whereas for conformal coupling, $\xi=1/6$, the corresponding coefficients are
\begin{equation}
\begin{aligned}
\gamma_{A,{\rm sc}}^{\rm c}
&=
\frac{
\left(
-2341\alpha^4+11078\alpha^2+32
\right)\Delta_\alpha
}{
136080\pi\mu_{\rm sc}^2
},
\\
\gamma_{B,{\rm sc}}^{\rm c}
&=
\frac{
\left(
13\alpha^4+13\alpha^2-8
\right)\Delta_\alpha
}{
34020\pi\mu_{\rm sc}^2
}.
\end{aligned}
\label{eq:scalar_conformal_gamma_definitions}
\end{equation}

The two independent local coefficients $\gamma_{A,{\rm sc}}$ and $\gamma_{B,{\rm sc}}$, defined in Eq.~\eqref{eq:scalar_gamma_definitions}, govern the respective $r^{-4}$ corrections to $A_{\rm sc}(r)$ and $B_{\rm sc}(r)$ in Eqs.~\eqref{eq:scalar_A_function} and \eqref{eq:scalar_B_function}. Their effect on the backreacted geometry is revealed by how these corrections modify the proper-time and proper-distance intervals associated with the two metric functions.

For a static observer, $dr=d\theta=d\varphi=0$, so the line element \eqref{eq:monopole_metric_ansatz} gives
\[
d\tau=\sqrt{A_{\rm sc}(r)}\,dt.
\]
Hence, for a fixed coordinate-time interval $dt$, the $r^{-4}$ term proportional to $\gamma_{A,{\rm sc}}$ changes the proper time accumulated by a clock held at radius $r$, and therefore changes the gravitational redshift between clocks at different radii. On the other hand, along a radial curve on a constant-time hypersurface, $dt=d\theta=d\varphi=0$, so that
\[
d\ell=\frac{dr}{\sqrt{B_{\rm sc}(r)}},
\]
which implies that the term proportional to $\gamma_{B,{\rm sc}}$ consequently changes the proper radial distance associated with a given coordinate separation $dr$.

Since Eqs.~\eqref{eq:scalar_A_function} and
\eqref{eq:scalar_B_function} were obtained as first-order corrections to
the classical background, their ratio can be expanded around
$A_{\rm sc}=B_{\rm sc}=\alpha^2$. The common term proportional to
$M_{\rm q,sc}/r$ cancels at this order, yielding
\[
\frac{A_{\rm sc}(r)}{B_{\rm sc}(r)}
\simeq
1
-
\frac{
\gamma_{A,{\rm sc}}
+
\gamma_{B,{\rm sc}}
}{
\alpha^2r^4}.
\]
Therefore, the sum $\gamma_{A,{\rm sc}}+\gamma_{B,{\rm sc}}$ determines the difference between the local scalar corrections to the temporal and radial metric functions. In general this sum is nonzero, so that $A_{\rm sc}(r)\neq B_{\rm sc}(r)$. This does not break spherical symmetry, because the angular part of the metric remains $r^2d\Omega^2$; it reflects instead the inequality $C_t^{\rm sc}\neq C_r^{\rm sc}$ between the temporal and radial components of the scalar stress-energy tensor.

The local terms proportional to $r^{-4}$ originate from the $r^{-6}$ behavior of the renormalized stress-energy tensor. The temporal Einstein equation first generates the $r^{-4}$ correction in $B_{\rm sc}(r)$. The difference between the radial and temporal Einstein equations then determines the additional contribution to $A_{\rm sc}(r)$ that is not shared by $B_{\rm sc}(r)$. Both local corrections are proportional to $\Delta_\alpha/\mu_{\rm sc}^2$. They are therefore suppressed for a heavier scalar field, as expected in the large-mass Schwinger--DeWitt expansion, and vanish in the limit $\Delta_\alpha\to0$, in which the monopole background becomes flat.

Using the proper-distance construction introduced in Sec.~\ref{sec:classical_monopole}, the scalar-corrected radial distance is obtained by replacing the classical radial metric function with $B_{\rm sc}(r)$. Equation~\eqref{eq:scalar_B_function} then gives
\begin{equation}
\ell_{\rm sc}(r)
\simeq
\frac{r-r_\star}{\alpha}
+
\frac{M_{\rm q,sc}}{\alpha^3}
\ln\!\left(\frac{r}{r_\star}\right)
+
\frac{\gamma_{B,{\rm sc}}}{6\alpha^3}
\left(
\frac{1}{r^3}
-
\frac{1}{r_\star^3}
\right).
\label{eq:scalar_proper_radius_asymptotics}
\end{equation}
The logarithmic term generated by $M_{\rm q,sc}$ and the inverse-power contribution generated by $\gamma_{B,{\rm sc}}$ are both subleading with respect to the term linear in $r$. Therefore, $\ell_{\rm sc}(r)/r$ approaches the same limit $1/\alpha$ as in the classical geometry.
As a consequence, since the angular sector is unchanged, the scalar-corrected spacetime has the same asymptotic solid angle, $4\pi\alpha^2$. Therefore, the quantum correction affects only the subleading relation between areal radius and proper radial distance, and produces a local deformation of the exterior geometry without modifying its asymptotic conical structure.

\section{Spinor-field backreaction}
\label{sec:spinor_backreaction}

We now specialize the general solution to a massive spinor field \cite{FernandezPiedra2020Spinor}. As established in Sec.~\ref{sec:renormalized_tensor}, its temporal and radial coefficients coincide, $C_t=C_r\equiv C_{\rm sp}$.
It then follows directly from Eqs.~\eqref{eq:A_general_solution} and \eqref{eq:B_general_solution} that the temporal and radial metric functions are equal,
\[
A_{\rm sp}(r)=B_{\rm sp}(r).
\]
Taking into account the spinor coefficient previously shown, and defining
\begin{equation}
\gamma_{\rm sp}
\equiv
\frac{8\pi}{3}C_{\rm sp}
=
\frac{
\Delta_\alpha
\left(
31\alpha^4+31\alpha^2+10
\right)
}{
15120\pi\mu_{\rm sp}^2
},
\label{eq:spinor_gamma_definition}
\end{equation}
we obtain for the metric functions 
\begin{equation}
A_{\rm sp}(r)
=
B_{\rm sp}(r)
=
\alpha^2
-
\frac{2M_{\rm q,sp}}{r}
-
\frac{\gamma_{\rm sp}}{r^4}.
\label{eq:spinor_metric_function}
\end{equation}
The equality $A_{\rm sp}(r)=B_{\rm sp}(r)$ does not imply that the spinor stress-energy tensor is isotropic, since its angular components remain different from its temporal and radial components. Instead, it follows directly from the temporal and radial semiclassical Einstein equations. Equivalently, the spinor geometry satisfies $g_{tt}g_{rr}=-1$.

As discussed in Sec.~\ref{sec:general_backreaction}, $M_{\rm q,sp}$ is an undetermined integration constant, whereas the coefficient $\gamma_{\rm sp}$ is fixed locally by the spinor vacuum polarization.

For a global monopole, $\Delta_\alpha>0$, and the polynomial $31\alpha^4+31\alpha^2+10$ is positive. Hence, $\gamma_{\rm sp}>0$.
When $M_{\rm q,sp}=0$, both $A_{\rm sp}(r)$ and $B_{\rm sp}(r)$ are smaller than their classical value $\alpha^2$ throughout the perturbative domain. The spinor vacuum polarization therefore produces a common local deformation of the temporal and radial sectors of the metric. Physically, since $\langle T^t{}_t \rangle \propto \gamma_{\rm sp} r^{-6} > 0$, the semiclassical energy density $\rho_q = -\langle T^t{}_t \rangle$ is strictly negative. This negative-energy vacuum polarization cloud surrounding the core implies that the effective mass enclosed within a given radius increases as one moves inward. Consequently, this produces an attractive local correction: a static observer must provide a stronger outward acceleration to remain at fixed radius compared to the classical background.

For $M_{\rm q,sp}=0$, Eq.~\eqref{eq:spinor_metric_function} gives
\[
A_{\rm sp}(r)
=
B_{\rm sp}(r)
=
\alpha^2
-
\frac{\gamma_{\rm sp}}{r^4}.
\]
Since the semiclassical solution was constructed as a first-order perturbation of the classical monopole metric, it is reliable only for $r\gg r_{\rm q,sp}$, where
\begin{equation}
r_{\rm q,sp}
=
\left(
\frac{\gamma_{\rm sp}}{\alpha^2}
\right)^{1/4}.
\label{eq:spinor_perturbative_scale}
\end{equation}
The radius $r_{\rm q,sp}$ is not a horizon radius, because the first-order approximation is no longer valid when $r$ is of this order.

The spinor solution is valid only in the common region where the pointlike monopole approximation requires $r\gg r_{\rm c}$, the large-mass Schwinger--DeWitt expansion requires $\mu_{\rm sp}r\gg\sqrt{\Delta_\alpha}$, and the perturbative treatment requires both $\left|2M_{\rm q,sp}/r\right|\ll\alpha^2$ and $\gamma_{\rm sp}/r^4\ll\alpha^2$.
When $M_{\rm q,sp}=0$, the last condition is equivalent to
$r\gg r_{\rm q,sp}$, with $r_{\rm q,sp}$ given in
Eq.~\eqref{eq:spinor_perturbative_scale}.

As in the scalar case, the spinor corrections are subleading at large radius. Therefore, they do not modify the asymptotic solid angle $4\pi\alpha^2$ and affect only the local exterior geometry.

\section{Vector-field backreaction}
\label{sec:vector_backreaction}

We now specialize the general solution to a massive vector field \cite{FernandezPiedra2020Vector}. As stated in Sec.~\ref{sec:renormalized_tensor}, the vector coefficient introduced in Eq.~\eqref{eq:vector_coefficient} satisfies
$C_t=C_r\equiv C_{\rm v}$. Therefore, as in the spinor case, the temporal and radial metric functions coincide,
$A_{\rm v}(r)=B_{\rm v}(r)$.

It is convenient to define
\begin{equation}
\gamma_{\rm v}
\equiv
\frac{8\pi}{3}C_{\rm v}
=
\frac{
\Delta_\alpha
\left(
25\alpha^4+25\alpha^2+4
\right)
}{
1260\pi\mu_{\rm v}^2
}.
\label{eq:vector_gamma_definition}
\end{equation}
Substitution into Eqs.~\eqref{eq:A_general_solution} and \eqref{eq:B_general_solution} gives
\begin{equation}
A_{\rm v}(r)
=
B_{\rm v}(r)
=
\alpha^2
-
\frac{2M_{\rm q,v}}{r}
-
\frac{\gamma_{\rm v}}{r^4}.
\label{eq:vector_metric_functions}
\end{equation}
Since $\Delta_\alpha>0$ and $25\alpha^4+25\alpha^2+4>0$, one has $\gamma_{\rm v}>0$. Therefore, the vector quantum correction decreases both metric functions in exactly the same qualitative manner as the spinor field, again reflecting a negative-energy vacuum polarization cloud. Consequently, the vector field's perturbative validity range and its preservation of the classical asymptotic solid angle follow identically the behavior discussed in Sec.~\ref{sec:spinor_backreaction}.

\section{Comparison between scalar, spinor, and vector backreactions}
\label{sec:comparison_spins}

The essential distinction between the three backreactions lies in the relation between the temporal and radial components of the quantum source. For a generic scalar coupling, these components generally differ, and the semiclassical correction modifies the temporal and radial sectors of the metric independently. Consequently, the scalar geometry is described by the two functions $A_{\rm sc}(r)$ and $B_{\rm sc}(r)$ in Eqs.~\eqref{eq:scalar_A_function} and \eqref{eq:scalar_B_function}. Physically, this means that the scalar vacuum polarization changes the redshift measured by static clocks and the proper radial distance in different ways.

In contrast, the spinor and vector sources satisfy $C_t=C_r$. Their temporal and radial metric functions therefore remain equal, as shown in Eqs.~\eqref{eq:spinor_metric_function} and \eqref{eq:vector_metric_functions}. 

To compare the quantum corrections produced by the spinor and vector fields, we focus on the coefficients of their $r^{-4}$ terms. The integration constants $M_{\rm q,sp}$ and $M_{\rm q,v}$ are not included, since they depend on the boundary conditions and on the matching to the monopole core rather than on the exterior quantum source. Using Eqs.~\eqref{eq:spinor_gamma_definition} and \eqref{eq:vector_gamma_definition}, we find
\begin{equation}
\frac{\gamma_{\rm v}}{\gamma_{\rm sp}}
=
12
\left(
\frac{\mu_{\rm sp}}{\mu_{\rm v}}
\right)^2
\frac{
25\alpha^4+25\alpha^2+4
}{
31\alpha^4+31\alpha^2+10
}.
\label{eq:comparison_vector_spinor_ratio_general_masses}
\end{equation}
This ratio determines which field produces the larger quantum deformation of the metric throughout the region in which both semiclassical solutions are valid.

For a fixed monopole parameter, the two corrections have equal magnitude when
\begin{equation}
\frac{\mu_{\rm v}}{\mu_{\rm sp}}
=
\sqrt{12\left(
\frac{
25\alpha^4+25\alpha^2+4
}{
31\alpha^4+31\alpha^2+10
}
\right)}.
\label{eq:comparison_vector_spinor_critical_mass_ratio}
\end{equation}
If the vector-to-spinor mass ratio is smaller than the value in Eq.~\eqref{eq:comparison_vector_spinor_critical_mass_ratio}, then $\gamma_{\rm v}>\gamma_{\rm sp}$ and the vector field produces the larger correction. If it is larger, the suppression associated with the larger vector mass compensates its larger numerical coefficient, and the spinor correction becomes dominant. Thus, the spin of the field fixes the numerical structure of the quantum source, while the masses determine the relative strength with which the two sources affect the geometry.

\begin{figure*}[t]
    \centering
    \includegraphics[width=0.72\textwidth]{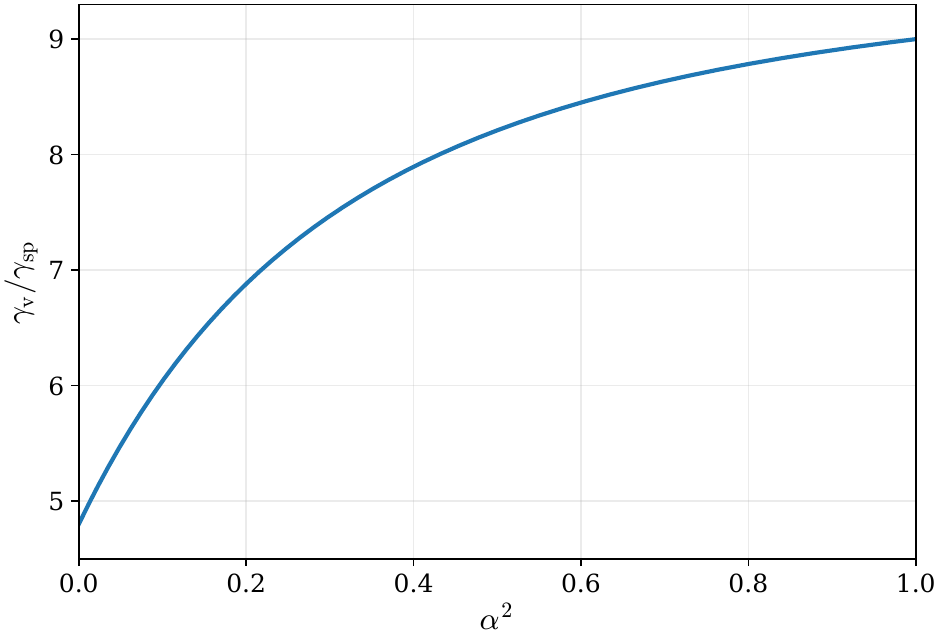}
    \caption{
    Ratio $\gamma_{\rm v}/\gamma_{\rm sp}$ for equal vector and spinor masses.
    The ratio increases monotonically with $\alpha^2$ throughout the physical
    interval $0<\alpha^2<1$. Hence, for equal masses, the vector field produces
    a larger locally fixed $r^{-4}$ correction than the spinor field.
    }
    \label{fig:vector-spinor-ratio}
\end{figure*}

The equal-mass case is obtained by setting $\mu_{\rm v}=\mu_{\rm sp}$ in Eq.~\eqref{eq:comparison_vector_spinor_ratio_general_masses}. The mass-dependent factor then becomes unity, so the relative strength of the two corrections is determined entirely by the monopole parameter $\alpha$. The resulting ratio increases monotonically throughout the physical interval $0<\alpha^2<1$: it approaches $24/5$ in the limit $\alpha^2\to0$ and tends to $9$ when $\alpha^2\to1$, as illustrated in Fig.~\ref{fig:vector-spinor-ratio}.

This increase describes a relative effect only. Both $\gamma_{\rm v}$ and $\gamma_{\rm sp}$ contain the common factor $\Delta_\alpha=1-\alpha^2$, and therefore both corrections vanish when $\alpha^2\to1$, where the monopole deficit and the background curvature disappear. Their ratio remains finite because this common factor cancels. Thus, in the small-deficit regime the vector correction is increasingly larger than the spinor correction relative to it, while the absolute magnitude of both quantum deformations simultaneously tends to zero.

For equal masses, the vector field therefore produces the larger locally fixed $r^{-4}$ correction at every radius within the common semiclassical domain. Since the spinor and vector geometries have the same radial dependence, this statement does not depend on the radius at which the two corrections are compared. It reflects only the different numerical coefficients that arise in the renormalized stress-energy tensors of fields with spin $1$ and spin $1/2$.

The comparison of the spinor and vector fields with the scalar one is more involved because the curvature coupling $\xi$ of the scalar field enters the quantum source through the two independent functions $\mathcal B(\alpha,\xi)$ and $\mathcal Q(\alpha,\xi)$. Consequently, the scalar correction is characterized by two coefficients, $\gamma_{A,{\rm sc}}$ and $\gamma_{B,{\rm sc}}$, which control the temporal and radial metric sectors separately.

For a generic value of $\xi$, the comparison with the vector field must therefore be performed independently in these two sectors. Whenever the corresponding scalar coefficient is nonzero, Eqs.~\eqref{eq:scalar_gamma_definitions} and \eqref{eq:vector_gamma_definition} give
\begin{align}
\frac{\gamma_{\rm v}}{\gamma_{A,{\rm sc}}}
&=
12
\left(
\frac{\mu_{\rm sc}}{\mu_{\rm v}}
\right)^2
\left(\frac{
25\alpha^4+25\alpha^2+4
}{
3\mathcal Q(\alpha,\xi)-\mathcal B(\alpha,\xi)
}\right),
\label{eq:comparison_vector_scalar_temporal_general}
\\
\frac{\gamma_{\rm v}}{\gamma_{B,{\rm sc}}}
&=
3
\left(
\frac{\mu_{\rm sc}}{\mu_{\rm v}}
\right)^2
\left(\frac{
25\alpha^4+25\alpha^2+4
}{
\mathcal B(\alpha,\xi)
}\right).
\label{eq:comparison_vector_scalar_radial_general}
\end{align}
The first ratio compares the local corrections to the temporal metric function, which controls the redshift between static observers. This correction also contributes to the acceleration required to remain at a fixed radius, together with the correction to the radial metric function. The second ratio compares the coefficients associated with the relation between coordinate radius and proper radial distance.

The signs of the scalar coefficients are physically important. The vector coefficient satisfies $\gamma_{\rm v}>0$, so the vector field decreases both $A_{\rm v}(r)$ and $B_{\rm v}(r)$ relative to their classical value when the core-dependent integration term is absent. For the scalar field, however, the signs of $\gamma_{A,{\rm sc}}$ and $\gamma_{B,{\rm sc}}$ depend on both $\alpha$ and $\xi$. If $\gamma_{A,{\rm sc}}>0$, the scalar and vector fields decrease the temporal metric function in the same direction; if $\gamma_{A,{\rm sc}}<0$, the scalar correction instead increases $A_{\rm sc}(r)$ and changes the redshift in the opposite direction.

When $\gamma_{B,{\rm sc}}>0$, the massive quantum vector field correction lowers $B_{\rm v}(r)$ and increases the proper radial distance associated with a fixed coordinate interval, whereas the quantum scalar field produces the opposite effect. If $\gamma_{B,{\rm sc}}<0$, both fields lower the radial metric function, although with generally different magnitudes.

The general scalar case also allows the temporal and radial responses to behave independently. For example, a value of $(\alpha,\xi)$ satisfying
$3\mathcal Q(\alpha,\xi)-\mathcal B(\alpha,\xi)=0$ gives $\gamma_{A,{\rm sc}}=0$, so that the locally fixed scalar correction to the temporal metric function vanishes while the radial correction remains nonzero unless $\mathcal B(\alpha,\xi)$ vanishes simultaneously. Conversely, $\mathcal B(\alpha,\xi)=0$ gives $\gamma_{B,{\rm sc}}=0$, while the temporal scalar correction can remain nonzero. Such possibilities have no analogue in the vector or spinor cases, for which the equality of temporal and radial source coefficients enforces a common metric function.

The corresponding comparison with the spinor field has the same two-sector structure. Although the spinor geometry satisfies $A_{\rm sp}(r)=B_{\rm sp}(r)$, the scalar field must again be compared with it separately through its temporal and radial coefficients. Using Eqs.~\eqref{eq:scalar_gamma_definitions} and \eqref{eq:spinor_gamma_definition}, one obtains
\begin{align}
\frac{\gamma_{\rm sp}}{\gamma_{A,{\rm sc}}}
&=
\left(
\frac{\mu_{\rm sc}}{\mu_{\rm sp}}
\right)^2
\left(\frac{
31\alpha^4+31\alpha^2+10
}{
3\mathcal Q(\alpha,\xi)-\mathcal B(\alpha,\xi)
}\right),
\label{eq:comparison_spinor_scalar_temporal_general}
\\
\frac{\gamma_{\rm sp}}{\gamma_{B,{\rm sc}}}
&=
\frac{1}{4}
\left(
\frac{\mu_{\rm sc}}{\mu_{\rm sp}}
\right)^2
\left(\frac{
31\alpha^4+31\alpha^2+10
}{
\mathcal B(\alpha,\xi)
}\right).
\label{eq:comparison_spinor_scalar_radial_general}
\end{align}
Since $\gamma_{\rm sp}>0$, the sign interpretation is the same as in the vector--scalar comparison. A positive $\gamma_{A,{\rm sc}}$ means that the scalar and spinor fields lower the temporal metric function in the same direction, whereas a negative $\gamma_{A,{\rm sc}}$ implies opposite corrections to the redshift sector. In the radial sector, a positive $\gamma_{B,{\rm sc}}$ means that the scalar and spinor corrections modify $B(r)$ with opposite signs, while a negative $\gamma_{B,{\rm sc}}$ makes both corrections decrease the radial metric function.

The mass ratios in Eqs.~\eqref{eq:comparison_spinor_scalar_temporal_general} and \eqref{eq:comparison_spinor_scalar_radial_general} express the inverse-square mass suppression of the Schwinger--DeWitt approximation. Thus, for fixed scalar coupling and monopole parameter, increasing the spinor mass decreases its contribution relative to the scalar one. Conversely, if the spinor is lighter, its one-function correction can dominate either scalar sector, depending on the values and signs of $\gamma_{A,{\rm sc}}$ and $\gamma_{B,{\rm sc}}$.

At scalar couplings for which $\gamma_{A,{\rm sc}}$ or $\gamma_{B,{\rm sc}}$ vanishes, the corresponding ratio diverges. This does not signal an enhancement of the spinor vacuum polarization. Rather, it reflects a cancellation within the scalar quantum source that removes one of the scalar metric corrections while the spinor correction remains nonzero. Such cancellations further illustrate that the scalar field possesses a degree of freedom absent from the spinor and vector cases: its temporal and radial backreactions can be independently suppressed, enhanced, or reversed by changing the curvature coupling.

For minimal coupling, $\xi=0$, the temporal scalar coefficient contains the same $\alpha$-dependent polynomial as the vector coefficient, whereas the spinor coefficient retains a different polynomial. The corresponding temporal ratios are
\begin{align}
\frac{\gamma_{\rm v}}{\gamma_{A,{\rm sc}}^{\rm m}}
&=
3
\left(
\frac{\mu_{\rm sc}}{\mu_{\rm v}}
\right)^2,
\label{eq:comparison_vector_scalar_minimal}
\\
\frac{\gamma_{\rm sp}}{\gamma_{A,{\rm sc}}^{\rm m}}
&=
\frac{1}{4}
\left(
\frac{\mu_{\rm sc}}{\mu_{\rm sp}}
\right)^2
\left(\frac{
31\alpha^4+31\alpha^2+10
}{
25\alpha^4+25\alpha^2+4
}\right).
\label{eq:comparison_spinor_scalar_minimal}
\end{align}
The vector--scalar ratio is independent of the monopole parameter. Therefore, for equal scalar and vector masses, the vector field produces a temporal $r^{-4}$ correction exactly three times larger than the minimally coupled scalar field. More generally, the vector contribution to the temporal sector is larger whenever $\mu_{\rm v}<\sqrt{3}\,\mu_{\rm sc}$, whereas a sufficiently heavier vector field is suppressed by the inverse-square mass dependence characteristic of the Schwinger--DeWitt expansion.

The spinor--scalar comparison is qualitatively different. Even for equal masses, the ratio in Eq.~\eqref{eq:comparison_spinor_scalar_minimal} depends on $\alpha$. It decreases monotonically from $5/8$ in the limit $\alpha^2\to0$ to $1/3$ as $\alpha^2\to1$. Hence, for equal masses, the minimally coupled scalar field always produces a larger temporal correction than the spinor field. The spinor contribution can exceed the scalar one only if the spinor is sufficiently lighter, namely when
\[
\frac{\mu_{\rm sp}}{\mu_{\rm sc}}
<
\sqrt{
\frac{
31\alpha^4+31\alpha^2+10
}{
4\left(25\alpha^4+25\alpha^2+4\right)
}}.
\]
This condition shows explicitly that the relative hierarchy is controlled jointly by the numerical coefficients associated with the field spin and by the inverse-square dependence on the masses.

The conformally coupled case, $\xi=1/6$, displays a richer dependence on the monopole parameter. Substitution of $\xi=1/6$ into the general scalar expressions gives
\begin{align*}
3\mathcal Q\left(\alpha,\frac16\right)
-
\mathcal B\left(\alpha,\frac16\right)
&=
\frac{
32+11078\alpha^2-2341\alpha^4
}{9},
\\
\mathcal B\left(\alpha,\frac16\right)
&=
\frac{
13\alpha^4+13\alpha^2-8
}{9}.
\end{align*}
The first of these quantities is positive throughout the physical interval $0<\alpha^2<1$. Thus, the conformally coupled scalar field always produces a temporal correction with the same sign as the vector and spinor fields. However, its magnitude depends strongly on $\alpha$:
\begin{align*}
\frac{\gamma_{\rm v}}{\gamma_{A,{\rm sc}}^{\rm c}}
&=
108
\left(
\frac{\mu_{\rm sc}}{\mu_{\rm v}}
\right)^2
\frac{
25\alpha^4+25\alpha^2+4
}{
32+11078\alpha^2-2341\alpha^4
},
\\
\frac{\gamma_{\rm sp}}{\gamma_{A,{\rm sc}}^{\rm c}}
&=
9
\left(
\frac{\mu_{\rm sc}}{\mu_{\rm sp}}
\right)^2
\frac{
31\alpha^4+31\alpha^2+10
}{
32+11078\alpha^2-2341\alpha^4
}.
\end{align*}
For equal masses, the equality between the vector and conformally coupled scalar temporal corrections, $\gamma_{\rm v}=\gamma_{A,{\rm sc}}^{\rm c}$ occurs for
\[
\alpha^2_{\rm v,sc}
=
\frac{59-\sqrt{3081}}{71}
\simeq0.0492.
\]
which is the only value that belongs to the physical interval $0<\alpha^2<1$
Likewise, the equality between the spinor and conformally coupled scalar temporal corrections, $\gamma_{\rm sp}=\gamma_{A,{\rm sc}}^{\rm c}$, requires the physical solution
\[
\alpha^2_{\rm sp,sc}
=
\frac{10799-\sqrt{116010561}}{5240}
\simeq0.00538.
\]
These two crossing values divide the equal-mass conformal case into three regimes. For $0<\alpha^2<\alpha^2_{\rm sp,sc}$, the temporal corrections satisfy $\gamma_{\rm v}
>
\gamma_{\rm sp}
>
\gamma_{A,{\rm sc}}^{\rm c}$. In the intermediate interval $\alpha^2_{\rm sp,sc}<\alpha^2<\alpha^2_{\rm v,sc}$, the conformally coupled scalar contribution becomes larger than the spinor one, while the vector contribution remains dominant $\gamma_{\rm v}
>
\gamma_{A,{\rm sc}}^{\rm c}
>
\gamma_{\rm sp}$. Finally, for $\alpha^2>\alpha^2_{\rm v,sc}$, the conformally coupled scalar field produces the largest temporal correction, $\gamma_{A,{\rm sc}}^{\rm c}
>
\gamma_{\rm v}
>
\gamma_{\rm sp}$. Hence, even when all three fields have the same mass, the hierarchy of the temporal backreaction changes as the monopole deficit is varied.

The radial sector of the conformally coupled scalar field behaves differently. Its coefficient vanishes when
\[
\alpha^2_{B,{\rm sc}}
=
\frac{\sqrt{585}-13}{26}
\simeq0.430.
\]
For $\alpha^2<\alpha^2_{B,{\rm sc}}$, one has $\gamma_{B,{\rm sc}}^{\rm c}<0$. The scalar contribution to $B_{\rm sc}(r)$ is then negative and therefore decreases the radial metric function in the same direction as the spinor and vector corrections. Consequently, all three fields increase the proper radial distance associated with a fixed coordinate interval.

For $\alpha^2>\alpha^2_{B,{\rm sc}}$, the sign changes to $\gamma_{B,{\rm sc}}^{\rm c}>0$. The conformally coupled scalar field then increases $B_{\rm sc}(r)$, whereas the spinor and vector fields decrease their radial metric functions. The scalar contribution consequently decreases the proper radial distance, in contrast with the spinor and vector contributions.

These results show that the scalar field cannot be assigned a single hierarchy relative to the spinor and vector fields. Even for the fixed conformal value $\xi=1/6$, its temporal correction changes its relative magnitude as the monopole parameter varies, while its radial correction undergoes an independent sign reversal. The spinor and vector fields do not display this separation because the equality of their temporal and radial source coefficients ties the two sectors of their backreaction together. The physical consequences of these differences are examined next through the curvature, the acceleration of static observers, and the motion of test particles.

\section{Physical implications and domain of validity of the semiclassical solutions}
\label{sec:physical_analysis}

The semiclassical metrics obtained above contain two distinct types of corrections. The terms proportional to $M_{\rm q}/r$ are integration constants whose values depend on the monopole core and on the boundary conditions used to connect the exterior solution with an interior geometry \cite{BezerraMelloSaharian2007}. By contrast, the $r^{-4}$ terms are fixed locally by the renormalized stress-energy tensor and describe the part of the backreaction generated directly by vacuum polarization. The aim of this section is to characterize this locally fixed geometry and to determine the radial region in which it can be interpreted as a small deformation of the classical global-monopole spacetime.

This distinction must be established before studying observables such as the acceleration of static observers or the motion of test particles. Those quantities can be calculated from the semiclassical metric at any radius, but their interpretation within the first-order approximation is reliable only where the quantum correction remains small compared with the classical metric. 

Throughout this section, we set $M_{\rm q,sc}=M_{\rm q,sp}=M_{\rm q,v}=0$. This does not impose a physical condition on the monopole core; it simply isolates the exterior contribution fixed by vacuum polarization. In particular, the $r^{-1}$ terms do not contribute to the Ricci scalar outside the core.

\subsection{Curvature correction and validity of the first-order geometry}
\label{subsec:curvature_validity}

The Ricci scalar provides an invariant description of the local curvature generated by the quantum source. For the pointlike global monopole background, the classical contribution is given by Eq.~\eqref{eq:R0}. Taking the trace of the semiclassical Einstein equations gives
\begin{equation}
R
=
R_0+\delta R
=
R_0
-
8\pi
\left\langle T^\mu{}_\mu\right\rangle_{\rm ren}^{(0)}.
\label{eq:Rtrace}
\end{equation}
Using the conserved quantum source in Eq.~\eqref{eq:conserved_quantum_tensor}, its trace is
\[
\left\langle T^\mu{}_\mu\right\rangle_{\rm ren}^{(0)}
=
\frac{1}{r^6}
\left(
C_t+C_r-2C_r-2C_r
\right)
=
\frac{C_t-3C_r}{r^6}.
\]
Therefore, the semiclassical Ricci scalar becomes
\begin{equation}
R(r)
=
\frac{2\Delta_\alpha}{r^2}
+
\frac{8\pi(3C_r-C_t)}{r^6}.
\label{eq:Rtotal}
\end{equation}
The first term is the curvature already present in the classical monopole spacetime, whereas the second is the local contribution generated by vacuum polarization. Since the quantum term falls as $r^{-6}$ while the classical monopole curvature falls as $r^{-2}$, the vacuum-polarization correction is concentrated toward the inner part of the exterior region. At sufficiently large radius, the local curvature approaches the classical monopole result.

Although Eq.~\eqref{eq:Rtotal} characterizes the geometric effect of the quantum field, the Ricci scalar alone does not determine the validity of the first-order metric solution. The metric was obtained by expanding around the classical monopole geometry, and its validity must therefore be tested directly from the size of the corrections to the metric functions. In particular, a cancellation in the trace of the quantum source can make the correction to $R$ small even when a correction to $A(r)$ or $B(r)$ is not negligible.

For a field of spin $s$, let $\delta A_s(r)$ and $\delta B_s(r)$ denote the locally fixed $r^{-4}$ contributions to the temporal and radial metric functions. Since the semiclassical solution was obtained by treating these terms as perturbations of the classical monopole metric, their relevance must be assessed relative to the classical value of the corresponding metric coefficient, namely $\alpha^2$. This motivates the dimensionless measures
\begin{equation}
\mathcal D_{A,s}(r)
\equiv
\frac{|\delta A_s(r)|}{\alpha^2},
\qquad
\mathcal D_{B,s}(r)
\equiv
\frac{|\delta B_s(r)|}{\alpha^2}.
\label{eq:relative_metric_corrections}
\end{equation}
that measure directly the fractional size of the quantum correction in the temporal and radial sectors. The absolute values are used because the validity of the perturbative expansion depends on the magnitude of the correction, independently of whether vacuum polarization increases or decreases the corresponding metric function. The first-order geometry is self-consistent only when both quantities are much smaller than unity.

For the scalar field, Eqs.~\eqref{eq:scalar_A_function} and \eqref{eq:scalar_B_function} give
\[
\mathcal D_{A,{\rm sc}}(r)
=
\frac{|\gamma_{A,{\rm sc}}|}{\alpha^2r^4},
\qquad
\mathcal D_{B,{\rm sc}}(r)
=
\frac{|\gamma_{B,{\rm sc}}|}{\alpha^2r^4}.
\]
The two quantities must be considered separately because the scalar stress-energy tensor has, in general, different temporal and radial components. Consequently, a small correction to the redshift function does not by itself guarantee that the radial metric function is also perturbatively close to its classical value.

It is useful to associate a radial scale with each scalar correction by asking at which radius its magnitude becomes comparable with $\alpha^2$. Since both corrections fall as $r^{-4}$, the corresponding scales are
\begin{equation}
r_{{\rm q},A}^{\rm sc}
=
\left(
\frac{|\gamma_{A,{\rm sc}}|}{\alpha^2}
\right)^{1/4},
\qquad
r_{{\rm q},B}^{\rm sc}
=
\left(
\frac{|\gamma_{B,{\rm sc}}|}{\alpha^2}
\right)^{1/4}.
\label{eq:scalar_metric_scales}
\end{equation}
The first-order scalar metric requires both sectors to remain perturbative.
Its common validity condition is therefore
$r\gg r_{\rm q}^{\rm sc}$, where
$r_{\rm q}^{\rm sc}\equiv\max(r_{{\rm q},A}^{\rm sc},
r_{{\rm q},B}^{\rm sc})$. These scales are reference scales for the
perturbative expansion, not physical surfaces. If either coefficient
vanishes for a particular value of $\alpha$ and $\xi$, the associated local
correction is absent.

For the spinor and vector fields, the equality of the temporal and radial
metric functions implies that a single local correction controls both sectors.
Accordingly,
\begin{equation}
\begin{aligned}
\mathcal D_{\rm sp}(r)
&\equiv
\mathcal D_{A,{\rm sp}}(r)
=
\mathcal D_{B,{\rm sp}}(r)
=
\frac{\gamma_{\rm sp}}{\alpha^2r^4},
\\
\mathcal D_{\rm v}(r)
&\equiv
\mathcal D_{A,{\rm v}}(r)
=
\mathcal D_{B,{\rm v}}(r)
=
\frac{\gamma_{\rm v}}{\alpha^2r^4}.
\end{aligned}
\label{eq:spinor_vector_metric_corrections}
\end{equation}
The associated scales are
\begin{equation}
r_{\rm q}^{\rm sp}
=
\left(
\frac{\gamma_{\rm sp}}{\alpha^2}
\right)^{1/4},
\qquad
r_{\rm q}^{\rm v}
=
\left(
\frac{\gamma_{\rm v}}{\alpha^2}
\right)^{1/4}.
\label{eq:spinor_vector_metric_scales}
\end{equation}
The spinor and vector geometries are perturbatively reliable when
$r\gg r_{\rm q}^{\rm sp}$ and $r\gg r_{\rm q}^{\rm v}$, respectively.
\begin{figure*}[t]
    \centering    \includegraphics[width=0.72\textwidth]{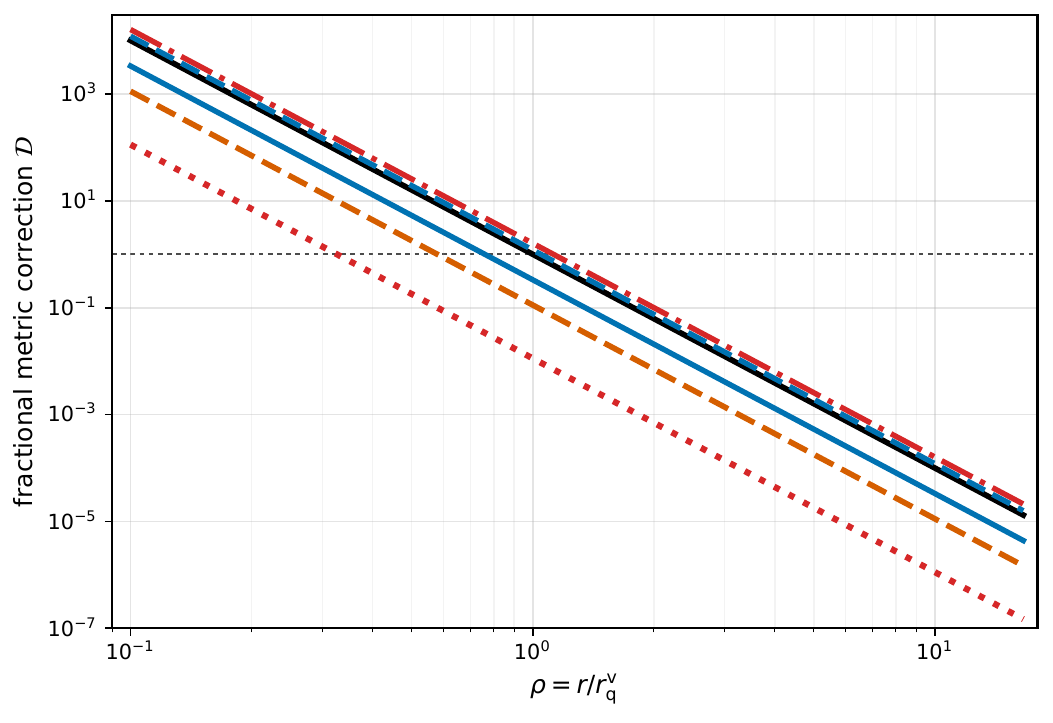}
    \caption{
Fractional locally fixed metric corrections as functions of
$\rho=r/r_{\rm q}^{\rm v}$ for $\alpha^2=0.90$ and equal field masses.
Both axes are logarithmic. The black solid curve represents the vector
correction $\mathcal D_{\rm v}$, while the orange dashed curve represents
the spinor correction $\mathcal D_{\rm sp}$. For the minimally coupled
scalar field, $\xi=0$, the blue solid and blue dashed curves represent,
respectively, $\mathcal D_{A,{\rm sc}}$ and $\mathcal D_{B,{\rm sc}}$.
For the conformally coupled scalar field, $\xi=1/6$, the red dash-dotted
and red dotted curves represent, respectively,
$\mathcal D_{A,{\rm sc}}$ and $\mathcal D_{B,{\rm sc}}$. The gray
short-dashed horizontal line marks $\mathcal D=1$, where the magnitude of
the corresponding locally fixed correction equals the classical metric
coefficient $\alpha^2$.
}    \label{fig:metric_correction_profiles}
\end{figure*}
Figure~\ref{fig:metric_correction_profiles} illustrates the direct metric
criterion for equal field masses at $\alpha^2=0.90$, or equivalently
$\Delta_\alpha=0.10$. This value is used as a representative finite-deficit
configuration: it is sufficiently far from the limiting cases
$\alpha^2\to0$ and $\alpha^2\to1$ to make the relative amplitudes of the
different corrections visible. Equal masses are assumed only to remove the
otherwise trivial hierarchy associated with the inverse-square mass
dependence and to isolate the effects of field spin and scalar curvature
coupling.

To display all sectors on a common dimensionless radial axis, we use
$\rho=r/r_{\rm q}^{\rm v}$. This is a graphical normalization, not a
physical preference for the vector field. The vector sector is convenient
because it has a single nonzero correction scale throughout the physical
interval. The scalar field instead has separate temporal and radial scales,
and either of them can vanish for particular values of $\alpha$ and $\xi$.
The spinor scale could equally be used; this would only rescale the horizontal
axis without changing the relative ordering or radial behavior of the
curves.

With this normalization, the vector correction satisfies
$\mathcal D_{\rm v}=1$ at $\rho=1$. This point is a reference scale rather
than part of the perturbative regime. Both axes are logarithmic because the
corrections span several orders of magnitude over the displayed radial
interval. The logarithmic representation also makes the common
$\rho^{-4}$ behavior appear as parallel straight lines, so that the
field-dependent amplitudes can be compared without obscuring the
large-radius region where the corrections become small.

At any fixed value of $\rho$, a curve lying higher in the plot corresponds
to a larger fractional correction to the associated metric function. At
$\rho=1$, the temporal conformal-scalar correction and the radial
minimal-scalar correction lie above the vector curve. Thus, at the vector
reference radius, these two contributions are still larger than the
classical metric coefficient and require a larger radius to become
perturbatively small. The temporal minimal-scalar correction, the spinor
correction, and the radial conformal-scalar correction lie below the vector
curve, so they become small at shorter radial scales. Since all curves are
proportional to $\rho^{-4}$, their ordering is unchanged throughout the
common validity region.

The exterior semiclassical solution is applicable only where three
independent requirements are simultaneously satisfied. First, the observation
point must lie outside the physical monopole core, so that the pointlike
description applies. Second, the field mass must be sufficiently large
relative to the local curvature scale for the Schwinger--DeWitt expansion to
be valid. Third, the locally fixed metric correction must remain small. These
conditions are
\begin{equation}
r
\gg
r_{\rm c},
\qquad
\mu_s r
\gg
\sqrt{\Delta_\alpha},
\qquad
r
\gg
r_{\rm q}^{s}.
\label{eq:validity}
\end{equation}
For the scalar field, $r_{\rm q}^{s}$ denotes $r_{\rm q}^{\rm sc}$. For the
spinor and vector fields, it denotes $r_{\rm q}^{\rm sp}$ and
$r_{\rm q}^{\rm v}$, respectively.

\begin{figure*}[t]
    \centering
    \includegraphics[width=0.72\textwidth]{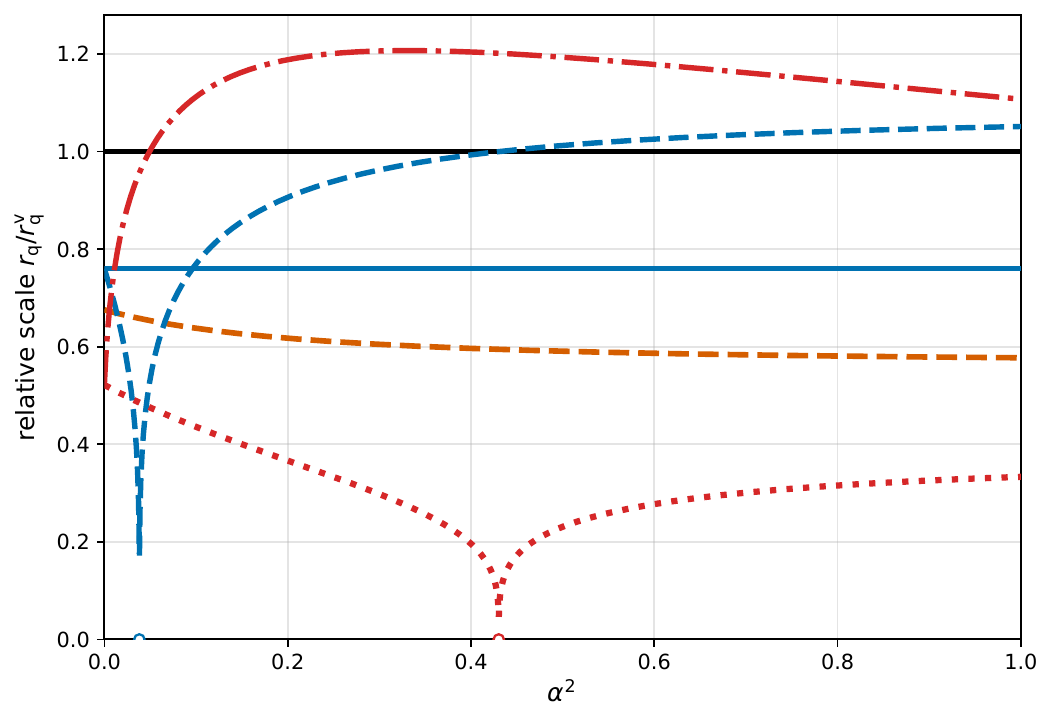}
   \caption{
Relative characteristic metric-correction scales for equal field masses,
normalized by $r_{\rm q}^{\rm v}$, as functions of $\alpha^2$. The black
solid line represents the vector ratio
$r_{\rm q}^{\rm v}/r_{\rm q}^{\rm v}=1$, and the orange dashed curve
represents $r_{\rm q}^{\rm sp}/r_{\rm q}^{\rm v}$. For the minimally
coupled scalar field, $\xi=0$, the blue solid and blue dashed curves
represent, respectively,
$r_{{\rm q},A}^{\rm sc}/r_{\rm q}^{\rm v}$ and
$r_{{\rm q},B}^{\rm sc}/r_{\rm q}^{\rm v}$. For the conformally coupled
scalar field, $\xi=1/6$, the red dash-dotted and red dotted curves
represent, respectively,
$r_{{\rm q},A}^{\rm sc}/r_{\rm q}^{\rm v}$ and
$r_{{\rm q},B}^{\rm sc}/r_{\rm q}^{\rm v}$. The open circles indicate
the values of $\alpha^2$ at which the corresponding radial scalar
coefficient vanishes and, consequently, the associated characteristic
scale is zero. A ratio greater than unity means that the corresponding
metric correction becomes perturbatively small only at radii larger than
the vector reference scale; a ratio smaller than unity means that this
occurs at smaller radii.
}\label{fig:metric_validity_scales}
\end{figure*}

Figure~\ref{fig:metric_validity_scales} complements
Fig.~\ref{fig:metric_correction_profiles}. Whereas the first figure fixes
$\alpha^2=0.90$ and shows how the corrections decrease with radius, the
second shows how the characteristic scales $r_{\rm q}$ vary as the monopole
parameter changes. The plotted quantities are the ratios
$r_{\rm q}/r_{\rm q}^{\rm v}$ for equal field masses. A value above unity
means that the corresponding correction remains of order $\alpha^2$ out to a
larger radius than the vector correction, whereas a value below unity means
that it becomes perturbatively small at a shorter radius.

The sharp descents of the two radial scalar curves have a physical origin.
For minimal coupling, the radial scalar correction vanishes at
$\alpha^2\simeq0.0381$, whereas for conformal coupling it vanishes at
$\alpha^2\simeq0.430$. At each of these values, the locally fixed
$r^{-4}$ contribution to $B_{\rm sc}(r)$ is absent. Consequently, the
associated characteristic scale vanishes, and the radial scalar sector does
not impose a perturbative lower radial bound there. These zeros reflect
cancellations in the scalar vacuum-polarization coefficient, rather than
numerical singularities.

\subsection{Acceleration of static observers}
\label{subsec:static_acceleration}

For the classical pointlike monopole with vanishing Schwarzschild-like mass
parameter, the worldlines with fixed $r$, $\theta$, and $\varphi$ require no
radial proper acceleration, since $A(r)=B(r)=\alpha^2$ is constant. The
semiclassical correction changes this property by introducing a radial
dependence in the temporal metric function.

For a static observer in the metric \eqref{eq:monopole_metric_ansatz}, the
four-velocity is
\begin{equation}
u^\mu
=
\left(
\frac{1}{\sqrt{A(r)}},
0,
0,
0
\right).
\label{eq:ustatic}
\end{equation}
Its radial four-acceleration is
\begin{equation}
a^r
=
u^\nu\nabla_\nu u^r
=
\frac{B(r)A'(r)}{2A(r)}.
\label{eq:arstatic}
\end{equation}
Since $a^r$ is a coordinate component, the locally measured radial
acceleration is more conveniently described by its component in the
outward-pointing orthonormal direction,
\begin{equation}
a^{\hat r}
\equiv
\frac{a^r}{\sqrt{B(r)}}
=
\frac{\sqrt{B(r)}\,A'(r)}{2A(r)}.
\label{eq:proper_static_acceleration}
\end{equation}
Its magnitude is the proper acceleration measured by the static observer,
$a_{\rm prop}=|a^{\hat r}|$. In the classical pointlike geometry,
$a^r=a^{\hat r}=0$.

After setting the core-dependent integration constants to zero, as in the
rest of this section, the general temporal metric function in
Eq.~\eqref{eq:A_general_solution} can be written as
\begin{equation}
A(r)
=
\alpha^2
-
\frac{\Gamma_A}{r^4},
\qquad
\Gamma_A
\equiv
2\pi
\left(
C_r+\frac{C_t}{3}
\right).
\label{eq:GammaA}
\end{equation}
Using this expression in Eqs.~\eqref{eq:arstatic} and
\eqref{eq:proper_static_acceleration}, and retaining only the first
semiclassical order, gives
\begin{equation}
a^r
\simeq
\frac{2\Gamma_A}{r^5},
\qquad
a^{\hat r}
\simeq
\frac{2\Gamma_A}{\alpha r^5}.
\label{eq:arquantum}
\end{equation}
Thus the locally measured acceleration falls as $r^{-5}$, one inverse power
of $r$ faster than the correction to the temporal metric function. The
radial correction to $B(r)$ does not contribute at this order: its effect in
Eq.~\eqref{eq:arstatic} multiplies $A'(r)$ and is therefore of second
semiclassical order.

For the scalar field, $\Gamma_A=\gamma_{A,{\rm sc}}$, and hence
\begin{equation}
a^{\hat r}_{\rm sc}
\simeq
\frac{2\gamma_{A,{\rm sc}}}{\alpha r^5}.
\label{eq:arscalar}
\end{equation}
The static acceleration is therefore controlled only by the temporal scalar
coefficient. In particular, the radial coefficient
$\gamma_{B,{\rm sc}}$ does not affect the leading acceleration of a static
observer. For both minimal and conformal coupling,
$\gamma_{A,{\rm sc}}^{\rm m}>0$ and
$\gamma_{A,{\rm sc}}^{\rm c}>0$ throughout the physical interval
$0<\alpha^2<1$. A static observer must consequently accelerate outward to
remain at fixed radius, which corresponds to an attractive local
semiclassical correction. For a generic curvature coupling, the sign is
instead determined by $\gamma_{A,{\rm sc}}$ itself.

For the spinor and vector fields, the local static accelerations are
\begin{equation}
a^{\hat r}_{\rm sp}
\simeq
\frac{2\gamma_{\rm sp}}{\alpha r^5},
\qquad
a^{\hat r}_{\rm v}
\simeq
\frac{2\gamma_{\rm v}}{\alpha r^5}.
\label{eq:arspinvec}
\end{equation}
Since $\gamma_{\rm sp}>0$ and $\gamma_{\rm v}>0$, both fields also produce
an attractive local correction. For equal masses, the vector coefficient is
larger than the spinor coefficient, so the outward acceleration required to
hold a static observer is correspondingly larger in the vector geometry.

\begin{figure*}[t]
    \centering
    \includegraphics[width=0.72\textwidth]{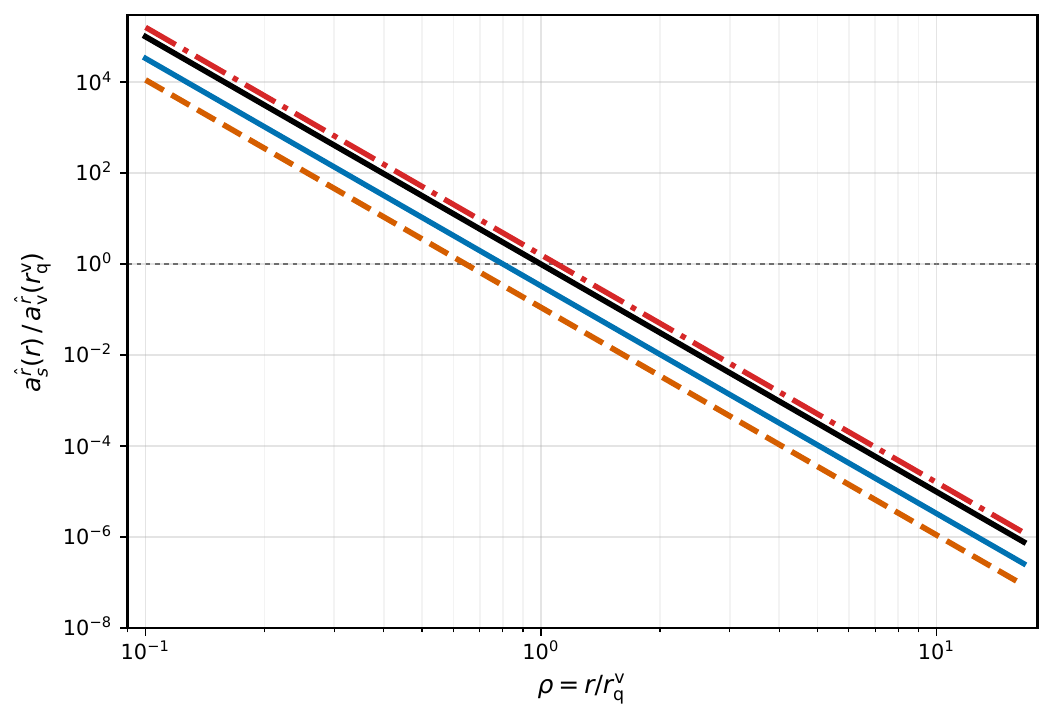}
    \caption{
    Normalized outward orthonormal acceleration required to keep a static
    observer at fixed radius, as a function of
    $\rho=r/r_{\rm q}^{\rm v}$, for $\alpha^2=0.90$ and equal field masses.
    Both axes are logarithmic. The black solid curve represents the vector
    field, the orange dashed curve the spinor field, the blue solid curve the
    minimally coupled scalar field, $\xi=0$, and the red dash-dotted curve
    the conformally coupled scalar field, $\xi=1/6$. The vertical normalization
    is $a^{\hat r}_{\rm v}(r_{\rm q}^{\rm v})$, so that the vector curve equals
    unity at $\rho=1$; the gray short-dashed horizontal line marks this
    reference level. The radial scalar coefficient does not enter the leading
    static acceleration, and consequently each scalar coupling is represented
    by a single curve.
    }
    \label{fig:static_acceleration_profiles}
\end{figure*}

To compare the four cases displayed below, we use the same radial
normalization as in Fig.~\ref{fig:metric_correction_profiles},
$\rho=r/r_{\rm q}^{\rm v}$, and define
\begin{equation}
\mathcal A_s(\rho)
\equiv
\frac{
a^{\hat r}_s(r)
}{
a^{\hat r}_{\rm v}(r_{\rm q}^{\rm v})
}.
\label{eq:normalized_static_acceleration}
\end{equation}
This is a graphical normalization only; $r_{\rm q}^{\rm v}$ remains a
reference scale rather than a radius at which the perturbative solution is
valid. Figure~\ref{fig:static_acceleration_profiles} shows that all four
accelerations decay as $\rho^{-5}$. For $\alpha^2=0.90$ and equal field
masses, the conformally coupled scalar field produces the largest outward
static acceleration, followed by the vector field, the minimally coupled
scalar field, and the spinor field. Only the part of each curve lying within
the common validity region specified by Eq.~\eqref{eq:validity} has a
semiclassical interpretation.

\subsection{Effective potential and radial kinetic factor}
\label{subsec:geodesics}

The radial dependence of the temporal metric function also modifies the
geodesic effective potential. By spherical symmetry, the motion can be
restricted to the equatorial plane. The conserved energy and angular momentum
are
\begin{equation}
E
=
A(r)\dot t,
\qquad
L
=
r^2\dot\varphi,
\label{eq:EL}
\end{equation}
where the dot denotes differentiation with respect to an affine parameter,
chosen as proper time for timelike geodesics. Using the normalization
condition for the four-velocity, the radial equation can be written as
\begin{equation}
\dot r^2
=
\frac{B(r)}{A(r)}
\left[
E^2
-
V_{\rm eff}^2(r)
\right],
\label{eq:radialgeo}
\end{equation}
with
\begin{equation}
V_{\rm eff}^2(r)
=
A(r)
\left(
\epsilon+\frac{L^2}{r^2}
\right),
\label{eq:Veff}
\end{equation}
where $\epsilon=1$ for timelike geodesics and $\epsilon=0$ for null
geodesics.

For the classical pointlike monopole background,
\begin{equation}
V_{{\rm eff},{\rm cl}}^2(r)
=
\alpha^2
\left(
\epsilon+\frac{L^2}{r^2}
\right).
\label{eq:Vcl}
\end{equation}
The locally fixed semiclassical correction modifies the effective potential
through the temporal metric function only. Combining
Eqs.~\eqref{eq:Veff} and \eqref{eq:Vcl}, one finds
\begin{equation}
\frac{
V_{\rm eff}^2(r)
-
V_{{\rm eff},{\rm cl}}^2(r)
}{
V_{{\rm eff},{\rm cl}}^2(r)
}
=
\frac{
A(r)-\alpha^2
}{
\alpha^2
}
=
-\frac{\Gamma_A}{\alpha^2r^4},
\label{eq:Vrel}
\end{equation}
where $\Gamma_A$ was defined in Eq.~\eqref{eq:GammaA}. Thus, the relative
potential correction is independent of whether the geodesic is timelike or
null, provided that $\epsilon+L^2/r^2\neq0$. For radial null geodesics,
$L=0$, the effective potential vanishes identically and the relative
quantity in Eq.~\eqref{eq:Vrel} is not defined.

For the scalar, spinor, and vector fields, respectively,
$\Gamma_A=\gamma_{A,{\rm sc}}$, $\gamma_{\rm sp}$, and $\gamma_{\rm v}$.
The spinor and vector coefficients are positive, so their locally fixed
corrections lower the effective potential. The same is true for the minimally
and conformally coupled scalar fields, for which
$\gamma_{A,{\rm sc}}^{\rm m}>0$ and
$\gamma_{A,{\rm sc}}^{\rm c}>0$ throughout the physical interval
$0<\alpha^2<1$. For a generic scalar curvature coupling, the sign of the
potential correction is determined directly by $\gamma_{A,{\rm sc}}$.

Equation~\eqref{eq:Vrel} is precisely the signed temporal metric correction.
Its magnitude is therefore already represented by the temporal curves in
Fig.~\ref{fig:metric_correction_profiles}; a separate plot of the relative
effective-potential correction would not provide independent information.

The scalar field produces an additional geodesic effect because its temporal
and radial metric functions are generally different. In contrast, the spinor
and vector geometries satisfy
$A_{\rm sp}=B_{\rm sp}$ and $A_{\rm v}=B_{\rm v}$, respectively, so that the
prefactor multiplying $E^2-V_{\rm eff}^2$ in Eq.~\eqref{eq:radialgeo} is
identically unity in those two cases.

Using Eqs.~\eqref{eq:scalar_A_function} and
\eqref{eq:scalar_B_function}, the scalar radial kinetic factor becomes
\begin{equation}
\frac{
B_{\rm sc}(r)
}{
A_{\rm sc}(r)
}
-
1
\simeq
\frac{
\gamma_{A,{\rm sc}}
+
\gamma_{B,{\rm sc}}
}{
\alpha^2r^4}.
\label{eq:BAdef}
\end{equation}
Consequently, the turning points of the radial motion are still determined
by $E^2=V_{\rm eff}^2(r)$, whereas the scalar prefactor changes the radial
evolution between those turning points.

For minimal coupling, Eq.~\eqref{eq:BAdef} gives
\begin{equation}
\left.
\left[
\frac{
B_{\rm sc}(r)
}{
A_{\rm sc}(r)
}
-
1
\right]
\right|_{\xi=0}
=
\frac{
\Delta_\alpha(1+\alpha^2)
}{
30\pi\mu_{\rm sc}^2r^4
}.
\label{eq:BAmin}
\end{equation}
For conformal coupling, one obtains
\begin{equation}
\left.
\left[
\frac{
B_{\rm sc}(r)
}{
A_{\rm sc}(r)
}
-
1
\right]
\right|_{\xi=1/6}
=
\frac{
\Delta_\alpha
\left(
3710-763\alpha^2
\right)
}{
45360\pi\mu_{\rm sc}^2r^4
}.
\label{eq:BAconf}
\end{equation}
Both expressions are positive throughout $0<\alpha^2<1$. Therefore, in the
classically allowed region, where $E^2>V_{\rm eff}^2(r)$, the scalar radial
kinetic prefactor increases $\dot r^2$ relative to the one-function spinor
and vector geometries with the same value of
$E^2-V_{\rm eff}^2(r)$.

\begin{figure*}[t]
    \centering
    \includegraphics[width=0.72\textwidth]{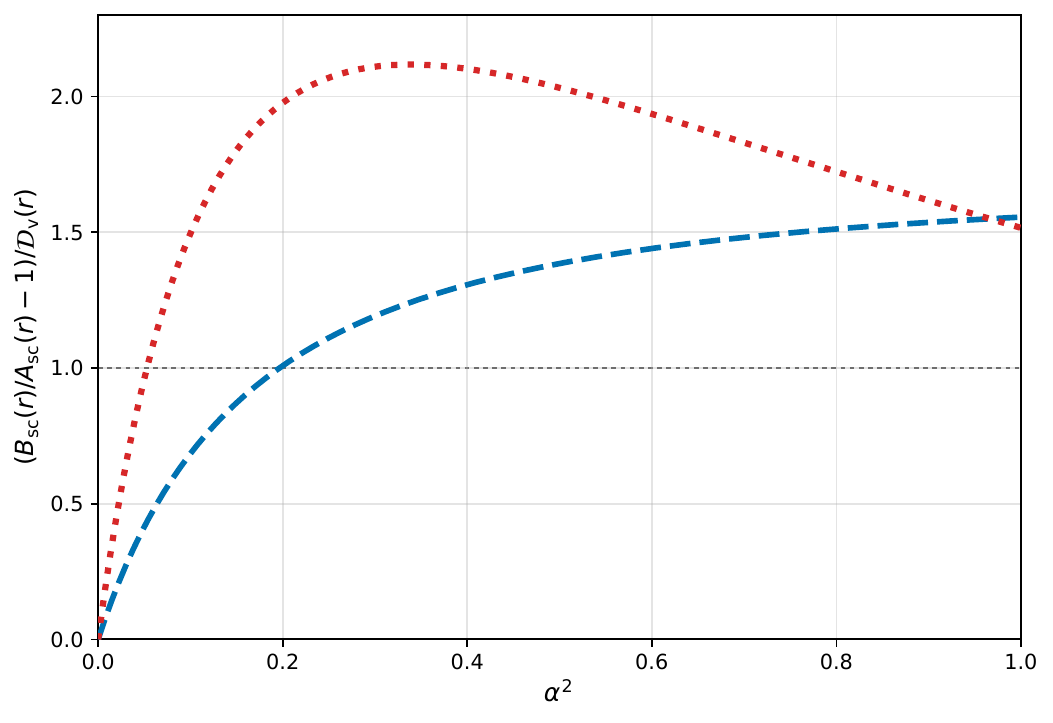}
    \caption{
    Relative strength of the scalar radial kinetic correction for equal
    scalar and vector masses. The plotted quantity is
    $\mathcal K_\xi=[B_{\rm sc}(r)/A_{\rm sc}(r)-1]/\mathcal D_{\rm v}(r)$,
    which is independent of radius at first semiclassical order. The blue
    dashed curve corresponds to minimal coupling, $\xi=0$, and the red dotted
    curve to conformal coupling, $\xi=1/6$. The gray short-dashed horizontal
    line marks $\mathcal K_\xi=1$, for which the scalar kinetic correction has
    the same magnitude as the vector fractional metric correction
    $\mathcal D_{\rm v}(r)$ at the same radius.
    }
    \label{fig:BAfig}
\end{figure*}

Since both Eqs.~\eqref{eq:BAmin} and \eqref{eq:BAconf} have the same
$r^{-4}$ dependence as $\mathcal D_{\rm v}(r)$, their relative strengths can
be compared without introducing an arbitrary field mass or radial reference
point. Figure~\ref{fig:BAfig} displays
\begin{equation}
\mathcal K_\xi(\alpha^2)
\equiv
\frac{
B_{\rm sc}(r)/A_{\rm sc}(r)-1
}{
\mathcal D_{\rm v}(r)
}
=
\frac{
\gamma_{A,{\rm sc}}+\gamma_{B,{\rm sc}}
}{
\gamma_{\rm v}
},
\label{eq:kinetic_prefactor_ratio}
\end{equation}
for equal scalar and vector masses. The ratio is independent of $r$ because
both numerator and denominator fall as $r^{-4}$. The conformally coupled
scalar field produces the larger radial kinetic correction over most of the
physical interval, whereas the two couplings become nearly equal in the
small-deficit regime.

The geodesic effects are therefore separated cleanly. The temporal metric
function determines the effective potential and its turning points, whereas
the scalar difference between $A_{\rm sc}(r)$ and $B_{\rm sc}(r)$ introduces
an additional radial kinetic factor. This second effect has no analogue in
the spinor and vector geometries considered here.

\section{Conclusions}
\label{sec:conclusions}

We have derived the leading semiclassical backreaction produced by massive
scalar, spinor, and vector fields in the spacetime of a pointlike global
monopole. Using the Schwinger--DeWitt approximation, we solved the
semiclassical Einstein equations for a general conserved quantum source and
then specialized the result to the three field types.

The backreacted metric contains a core-dependent $r^{-1}$ integration term
and a locally determined $r^{-4}$ contribution generated by vacuum
polarization. The latter is the universal exterior correction studied here.
It is suppressed by the inverse square of the field mass and decreases
rapidly with radius, so it does not alter the asymptotic solid-angle deficit
of the classical monopole.

The scalar field gives the most general geometry because its temporal and
radial quantum stress-energy components need not coincide. Its backreaction
therefore modifies separately the temporal and radial metric sectors, which
produces both a correction to the geodesic effective potential and an
additional radial kinetic factor. By contrast, the spinor and vector fields
lead to one-function geometries. Their locally fixed coefficients are
positive throughout the physical interval, implying an attractive local
correction: a static observer must accelerate outward to remain at fixed
radius. For equal masses, the vector field produces the larger correction of
the two one-function cases.

All results apply only in the common region where the pointlike monopole
description, the large-mass Schwinger--DeWitt expansion, and the
first-order semiclassical treatment are simultaneously valid. Within this
domain, the solutions provide an analytic description of how field spin and
scalar curvature coupling determine the local quantum deformation of the
monopole exterior.

A natural next step is to replace the idealized pointlike monopole by a
regular finite-core configuration. Matching the exterior semiclassical
solution to an interior geometry would determine the integration constants
that remain undetermined by the local exterior equations and would establish
how the vacuum-polarization correction connects to the core region. Such an
analysis would also make it possible to assess quantitatively the overlap
between the finite-core description, the large-mass approximation, and the
perturbative semiclassical regime.

It would also be useful to extend the calculation beyond the leading
Schwinger--DeWitt term and to compare the resulting local approximation with
renormalized stress-energy tensors obtained by numerical methods \cite{FernandezPiedra2020Phi2,AndersonHiscockSamuel1995,
MatyjasekTrynieckiZwierzchowska2010}. This would
clarify the accuracy of the leading large-mass result as the observation
radius approaches the region where the local curvature becomes larger.
Finally, the same framework can be applied to global-monopole geometries
with a nonzero Schwarzschild-like mass parameter or with horizons. In those
cases, the locally fixed $r^{-4}$ terms derived here would coexist with
additional classical scales, allowing one to investigate how vacuum
polarization modifies the near-horizon geometry, the static-observer
acceleration, and the geodesic structure \cite{AndersonHiscockTaylor2000}.

\begin{acknowledgments}
The author acknowledges financial support from the Fundação de Apoio à
Pesquisa do Estado da Paraíba (FAPESQ).
\end{acknowledgments}

\bibliographystyle{apsrev4-2}
\bibliography{references_global_monopole}

\end{document}